\documentstyle[12pt]{article}
\begin{document}

\vspace{1.0cm}

\begin{center}
{\bf MODIFICATION OF BLACK-HOLE ENTROPY BY STRINGS}
\end{center}

\vspace{1.0cm}

\begin{center}
R.Parthasarathy{\footnote{Permanent Address: The Institute of
Mathematical Sciences, C.I.T.Campus, Taramani Post, Madras 600113,India;
e-mail: rparthas@sfu.ca and sarathy@imsc.ernet.in}}
and K.S.Viswanathan{\footnote{e-mail: kviswana@sfu.ca}} \\
Department of Physics \\
Simon Fraser University \\
Burnaby, B.C., Canada V5A 1S6. \\
\end{center}

\vspace{1.0cm}

\noindent{\it {ABSTRACT}}

A generalized action for strings which is a sum of the Nambu-Goto and the
extrinsic curvature (the energy integral of the surface) terms, is used
to couple strings to gravity. It is shown that the conical singularity
has deficit angle that has contributions from both the above terms. It is
found that the effect of the extrinsic curvature is to oppose that of the
N-G action for the temperature of the black-hole and to modify the
entropy-area relation.
\newpage

\vspace{0.5cm}

Recently Englert, Houart and Windey [1] obtained a relation between the
entropy and the area of a black-hole that is different from the
Bekenstein-Hawking [2,3] relation $ S = A/4$ ($G = 1$ units) when a
conical singularity producing source is present in the euclidean section
at $ r = 2M$. It is known that a conical singularity at $r = 2M$
modifies the euclidean periodicity of the blackhole metric and hence the
temperature [4,5,6]. It is necessary to take into account the source
producing the singularity to maintain euclidean saddle point. The authors
in Ref.1, introduced for this purpose an elementary string in the action.
The conical singularity arises from the string when it wraps around the
horizon and the resulting
deficit angle is determined by the string tension. In the evaluation of
the free energy, the contribution of the string term exactly cancels that
of the Einstein term and so only the boundary terms contribute. As a
result one obtains,
\begin{eqnarray}
S & =& \frac{A}{4} (1 - 4\mu)
\end{eqnarray}
where $S$ is the entropy of the blackhole, $A$ its horizon area and $\mu
$ the string tension. The string action
considered in [1] is the Nambu-Goto action,
\begin{eqnarray}
I_{string} & = & -\mu \int \sqrt{-g} d\sigma d\tau ,
\end{eqnarray}
where $g$ is the determinant of the induced metric on the worldsheet.

\vspace{0.5cm}

This observation i.e. (1), is important in the light of the recent
statistical derivation of Bekenstein-Hawking $S = A/4$ relation in
string theory by counting the (microscopic) BPS bound state degeneracy
[7,8,9,10,11]. As a result of introducing (2) on the horizon, the
temperature of the blackhole is increased as $T = T_H/(1 - 4\mu)$ leading
to an acceleration of the evaporation.

\vspace{0.5cm}

The key observation in [1] is the existence of non-trivial solution to
the string equations of motion from (2) when the string wraps around the
euclidean continuation of the horizon, a sphere at $r = 2M$. When the
string worldsheet wraps around the sphere, it has non-zero {\it
{extrinsic curvature}} and so the simple string action (2) needs a
generalization. The Nambu-Goto action is just the area term while the
'energy of the surface' is given by an action involving extrinsic
curvature as $\int \sqrt{-g}{|H|}^2 d\sigma d\tau$. It will be shown that
the space-time energy momentum tensor for the Nambu-Goto action
(see the first line
in (12)) has the property that there is no flux normal to the surface. On
the other hand, the energy momentum tensor for the extrinsic curvature
has a non-vanishing flux normal to the surface. Consequently it is
conceivable that this flux flow might tend to slow down the acceleration
of the evaporation, favouring the stability of the black-hole against the
effect in [1]. Indeed our present study confirms this.

\vspace{0.5cm}

The generalized string action [12,13,14] we use in this
article is a sum of the N-G action and the one involving extrinsic
geometry and is given by,
\begin{eqnarray}
I_{string} &=& -\mu \int \sqrt{-g} d\sigma d\tau + \frac{1}{{\alpha}_0^2}
\int \sqrt{-g} {|H|}^2 d\sigma d\tau ,
\end{eqnarray}
where ${|H|}^2\ =\ \sum_i \ H^iH^i$, $H^i\ =\ \frac{1}{2}\
H^{i\alpha\beta}\ g_{\alpha\beta}$, $H^{i\alpha\beta}$ is the second
fundamental form and i runs from $1$ to $D-2$. The string
worldsheet is considered as a 2-dimensional surface immersed in a
D-dimensional space-time. $H^{i\alpha\beta}$ is then defined by the Gauss
equation
\begin{eqnarray}
{\partial}_{\alpha}{\partial}_{\beta}X^{\mu} +
{\tilde{\Gamma}}^{\mu}_{\nu\rho}{\partial}_{\alpha}X^{\nu}{\partial}_{\beta}
X^{\rho} - {\Gamma}^{\gamma}_{\alpha\beta}{\partial}_{\gamma}X^{\mu} &=&
H^{i}_{\alpha\beta} N^{i\mu},
\end{eqnarray}
where $X^{\mu}(\sigma,\tau)$ are the immersion coordinates of the string
worldsheet, $\tilde{\Gamma}$ is the affine connection for the (curved)
D-dimensional space-time, $\Gamma$ is the affine connection on the string
worldsheet, calculated using the induced metric
\begin{eqnarray}
g_{\alpha\beta} &=& {\partial}_{\alpha}X^{\mu} {\partial}_{\beta}X^{\nu}
h_{\mu\nu},
\end{eqnarray}
$h_{\mu\nu}$ being the metric on the D-dimensional space-time, $N^{i\mu}$
are the $(D-2)$ normals to the string worldsheet and ${\alpha}^2_0$ is a
dimensionless coupling constant.

\vspace{0.5cm}

The present authors have earlier studied [15,16,17,18] the extrinsic
geometry of strings in some detail including instanton effects in
constant mean curvature ($H \neq 0$) and minimal ($H =0$) surfaces
[18] and recently [19]
considered the intrinsic and extrinsic geometrc properties of the string
worldsheets in curved space-time background. In this formalism the only
string dynamical degrees of freedom are its immersion coordinates
$X^{\mu}(\sigma,\tau)$. It is the purpose of this article to consider the
effect of introducing the generalized string action (3) in place of (2)
when the string wraps around the euclidean horizon of a Schwarzschild
black hole. We find the entropy-area relation gets modified as
\begin{eqnarray}
S &=& \frac{A}{4} \frac{(1 - 4\mu + ({\alpha}_0 M)^{-2})^2}{(1 - 4\mu -
({\alpha}_0 M)^{-2})}.
\end{eqnarray}
The temperature of the black-hole is found to be \newline
$T = 8\pi {\{M(1-4\mu) +
({\alpha}_0^2M)^{-1}\}}^{-1}$. This feature of $M$ dependence for $T$ is
shared by quantum one-loop effects [20].

\vspace{0.5cm}

The Lorentzian action for gravity coupled to matter fields is
\begin{eqnarray}
I&=&\frac{1}{16\pi}\int \sqrt{-h} R d^4X - \frac{1}{8\pi}\int_{\Sigma}
\sqrt{-h'}K d^3X   \nonumber \\
&+& \frac{1}{8\pi}\int_{{\Sigma}^0} \sqrt{-h^0}K^0 d^3X
+ I_{m},
\end{eqnarray}
where the first term is the usual Einstein-Hilbert action, $K$ is the
trace of the extrinsic curvature of the boundary $\Sigma $ of space-time,
$K^0$ is the same as $K$ but of the boundary ${\Sigma}^0$ of flat
space-time and $I_{m}$ is given in (3). The boundary
terms in (7) are first introduced by Gibbons and Hawking [21] and their
roles in cancelling the boundary terms arising from the Einstein-Hilbert
action and removing the divergence at space-like infinity due to the
K-term for asymptotically flat space-times are given in [22]. It is to be
noted here that we have two extrinsic curvatures; one is the trace of the
extrinsic curvature of the boundary of the space-time (denoted by $K$ in
(7)) and the other that of the 2-dimensional surface (the string world
sheet) immersed in the space-time (denoted by $H$ in (3))
which is not a boundary term. The above
action (7) admits the usual Schwarzschild blackhole solution corresponding to
trivial solution for $I_{m}$ i.e., zero string area and vanishing
extrinsic curvature. The Euclidean continuation of the Schwarzschild
solution is
\begin{eqnarray}
ds^2 &=& (1 - \frac{2M}{r})d{\tau}^2 + (1 - \frac{2M}{r})^{-1}dr^2 + r^2
d{\Omega}^2.
\end{eqnarray}
Following Hawking [23], we consider an euclidean section via $x\ =\ 4M(1
- \frac{2M}{r})^{\frac{1}{2}}$, $x\ \geq \ 0$, $0\ \leq \ \tau \ \leq \
8\pi M$. The periodicity of $\tau$ is $8\pi M$ and this gives,
\begin{eqnarray}
{\beta}_H &=& 8\pi M.
\end{eqnarray}
In the euclidean formalism of blackholes, the topology of the space-time
is $R^2 \times  S^2$ and with the angular variable $\tau $, the
topology becomes $D^2 \times  S^2$. There is no cusp now and the
deficit angle is zero. The euclidean action following from (7) is
\begin{eqnarray}
I_E &=& -\frac{1}{16\pi}\int \sqrt{h}R d^4X + \frac{1}{8\pi}
\int_{\Sigma} \sqrt{h'}K d^3X - \frac{1}{8\pi}
\int_{{\Sigma}^0}\sqrt{{h}^0}K^0 d^3X \nonumber \\
 &+& \mu \int \sqrt{g} d^2\sigma - \frac{1}{{\alpha}^2_0}\int \sqrt{g}
 {|H|}^2 d^2\sigma .
\end{eqnarray}
Variation of (10) with respect to the background metric $h_{\mu\nu}$
gives the Einstein equation
\begin{eqnarray}
R_{\mu\nu} - \frac{1}{2} h_{\mu\nu}R &=& -8\pi T_{\mu\nu},
\end{eqnarray}
where the energy momentum tensor $T_{\mu\nu}$ for the generalized string
action (3) is given by [19]
\newcommand{\Dell}{\mbox{{$\frac{{\delta}^4(X -
X(\sigma,\tau))}{\sqrt{h(X(\sigma,\tau))}}$}}}
\begin{eqnarray}
T^{\mu\nu} &=& \mu \int d^2\sigma \Dell
 \sqrt{g} g^{\alpha\beta}
{\partial}_{\alpha}X^{\mu} {\partial}_{\beta}X^{\nu} \nonumber \\
&-&\frac{1}{{\alpha}^2_0}[\int d^2\sigma  \sqrt{g}
\Dell [ {|H|}^2 g^{\alpha\beta}
{\partial}_{\alpha}X^{\mu} {\partial}_{\beta}X^{\nu} \nonumber \\
&-& g^{\alpha\beta} ({\nabla}_{\beta}H^i)
(N^{i\nu}{\partial}_{\alpha}X^{\mu} + N^{i\mu}
{\partial}_{\alpha}X^{\nu})] \nonumber \\
&+& {\nabla}_{\rho} \int d^2\sigma \sqrt{g} \Dell g^{\alpha\beta}
{\partial}_{\alpha}X^{\mu} {\partial}_{\beta}X^{\nu} H^i N^{i\rho}]
\end{eqnarray}
In (12) ${\nabla}_{\alpha}$ is the covariant derivative with respect to
the induced metric $g_{\alpha\beta}$ on the worldsheet and
${\nabla}_{\rho}$ is with respect to the background space-time metric
$h_{\mu\nu}$. Note that $X$ in (12) is just the a space-time point,
whereas $X(\sigma,\tau)$ stands for the string worldsheet dynamical
variables. One sees from the 4-dimensional Dirac delta function in (12)
that $T^{\mu\nu}(X)$ vanishes unless $X$ is exactly on the string
worldsheet. Further, the flux of $T^{\mu\nu}_{N-G}$ , given by the first
line in (12) along the normal direction, $T^{\mu\nu}_{N-G} N^i_{\nu}(X)$
is zero. This is no longer true for that part of $T^{\mu\nu}$ coming from
the extrinsic curvature action as can be seen from (12).
 Variation of (10) with respect to the worldsheet
coordinates, $X^{\mu}(\sigma,\tau) \ \rightarrow \
X^{\mu}(\sigma,\tau) + \delta X^{\mu}(\sigma,\tau)$ has been evaluated in
[19] and by writing
\begin{eqnarray}
\delta X^{\mu} &=& {\xi}_j N^{j\mu} + {\xi}^{\alpha}
{\partial}_{\alpha}X^{\mu},
\end{eqnarray}
as normal and tangential variations, it has been shown that the equation
of motion for normal variation is
\begin{eqnarray}
&{\nabla}^2 H^i - 2H^i (H^2 + k) + H^jH^{j\alpha\beta}H^i_{\alpha\beta}&
\nonumber \\
&-g^{\alpha\beta}H^j
{\tilde{R}}_{\mu\nu\rho\sigma}{\partial}_{\alpha}X^{\nu}{\partial}_
{\beta}X^{\rho}N^{j\mu} N^{i\sigma}\ =\ 0, &
\end{eqnarray}
where $k\ =\ {\alpha}^2_0/\mu$, $i\ =\ 1,2$,
${\tilde{R}}_{\mu\nu\rho\sigma}$ is the Ricci tensor of the curved
background space. The tangential variation has been shown to be just the
contracted structure equation of Codazzi. It is clear from (14) that the
string worldsheet is non-trivial on the euclidean section alluded to above,
as it can admit solutions $H^i\ \neq \ 0$.
Taking the trace of (11) and integrating over $d^4X$ we find,
\begin{eqnarray}
\int d^4X \sqrt{h} R(X) &=& 16\pi \{ \mu A - \frac{1}{{\alpha}^2_0}\int
d^2\sigma \sqrt{g} {|H|}^2 \},
\end{eqnarray}
where $A$ is the area of the string worldsheet. The above result is
general in the sense that we have not restricted the string worlsheet to
have constant mean curvature.

\vspace{0.5cm}

When the string wraps around the euclidean horizon, a sphere at $r =
2M$, $R \neq  0$ and (14) can have non-trivial solution. In such a
case, the integral on the left hand side of (15) gets contribution from
the singularity at $r = 2M$. Following [1] and [6], we consider an
infinitesimal tubular neighbourhood $D^2 \times S^2$ of $r\ =\ 2M$.Then
\begin{eqnarray}
\int d^4X \sqrt{h}\ R(X) &=& A \int_{D^2} \sqrt{^{(2)}h}\ ^{(2)}R\ d^2X,
\end{eqnarray}
a product of $A$, the area of $S^2$, (identified as the horizon area) and
the Euler characteristic of disk. From (15) and (16) it follows that
\begin{eqnarray}
\frac{1}{4\pi} \int_{D^2} \sqrt{^{(2)}h}\ ^{(2)}R\ d^2X &=& 4\{ \mu -
\frac{1}{{\alpha}^2_0 A} \int d^2\sigma \sqrt{g} {|H|}^2 \}.
\end{eqnarray}
But, by Gauss-Bonnet theorem, the left hand side of (17) is the Euler
characteristic $\chi $ of the disk which is 1. Thus the introduction of
generalized string on the horizon has changed the topology of the disk by
creating a cusp. The deficit angle is
\begin{eqnarray}
\eta &=& 4\{ \mu - \frac{1}{{\alpha}^2_0 A} \int d^2\sigma \sqrt{g}
{|H|}^2 \}
\end{eqnarray}
Comparison with Eqn.16 of [1] shows that the generalized string with
extrinsic curvature has a non-trivial effect. For a sphere the mean
curvature is a constant [24], ${|H|}^2
= 1/r^2$. The effect of adding the extrinsic curvature action is thus to
modify the periodicity of $\tau$ from ${\beta}_H\ =\ 8\pi M$ to
\begin{eqnarray}
\beta &=& {\beta}_H\{ 1 - 4\mu + ({\alpha}_0M)^{-2}\},
\end{eqnarray}
where $r =  2M$ is used. It follows from (19) that in the presence of the
extrinsic curvature action, the increase in the temperature of the
black-hole due to the N-G action is reduced so that the acceleration of
evaporation is slowed down.

\vspace{0.5cm}

The calculation of the free energy of the black-hole is made a lot easier
in view of (15). Accordingly the contribution from the Einstein term
$\int \sqrt{h} R d^4X$ in (10) is exactly cancelled by the string
contribution including the action for the extrinsic curvature. Thus the
only contribution to the free energy comes from the boundary terms in
(10). This contibution has been evaluated by Gibbons and Hawking [21]
(see also [1]) as
\begin{eqnarray}
I_{Boundary} &=& \frac{{\beta}^2}{16\pi},
\end{eqnarray}
from which the free energy (${\beta}^{-1} \ I_{Boundary}$) is
\begin{eqnarray}
E_{free} &=& \frac{M}{2}.
\end{eqnarray}
The free energy is the same without the generalized string action. From
(19), using (9) we find,
\begin{eqnarray}
\frac{dM}{d\beta} &=& {\{ 8\pi (1 - 4\mu - ({\alpha}^2_0 M^2)^{-1}\}}^{-1},
\end{eqnarray}
and using the relation for entropy $S \ =\ {\beta}^2
\frac{dE_{free}}{d\beta}$ we have
\begin{eqnarray}
S &=& \frac{A}{4} \frac{{\{1 - 4\mu + ({\alpha}_0M)^{-2}\}}^2}{\{1 - 4\mu
- ({\alpha}_0M)^{-2}\}}.
\end{eqnarray}
Thus the in the presence of string wrapping the event horizon, the
entropy differs from that of the Bekenstein-Hawking $A/4$ relation.

\vspace{0.5cm}

Let us summarize our results.
We have argued that when the string worldsheet wraps around
the horizon, a 2-sphere ($r = 2M$), it has extrinsic curvature and this has
been added to the Nambu-Goto term in the form of 'energy integral' of the
surface.
The Einstein equations of motion with the
modified energy momentum tensor give the relation (15) from which it
follows that the free energy is independent of the string contribution as
it is cancelled by the Einstein term. By considering infinitesimal
tubular neighbourhood of $r = 2M$, the conical singularity has a deficit
angle (18). It is to be noted that the deficit angle depends on the
black-hole mass in contrast to the situation in [1]. This changes the
periodicity of $\tau$ and change the entropy-area relation according to
(23). When the mean curvature is constant, it has been shown [18] that
the string admits instanton solutions. Its effect in the context of QCD
was studied in [18]. In this article  another effect of string
instanton is described. In the absence of extrinsic curvature, the string
instanton has the effect of raising the global temperature of the
black-hole. The extrinsic curvature action has the opposite effect as far
as the temperature is concerned (19). It is intresting to note that the
inverse temperature $\beta$ of the black-hole has $M$ dependence through
the N-G action and $M^{-1}$ dependence through the extrinsic curvature
action. In [20] a similar feature was obtained from the one-loop quantum
effects, $M$ from the classical action and $M^{-1}$ from the quantum
correction. The extrinsic curvature
action can arise as quantum correction
when fermions on the string world sheet are functionally
integrated [25]. It can be
seen that the sign of $\sigma$ in [20] for spin-half fields is negative
agreeing with our ${\alpha}^2_0 > 0$.
 The entropy-area relation
is modified. The black-hole entropy can be greater that
$A/4$. It is possible to interpret (23) within the $S\ =\ A/4$ relation
if we introduce an effective area for the horizon as
\begin{eqnarray}
A_{eff} &=& A{\{1 - 4\mu + ({\alpha}_0M)^{-2}\}}^2 {\{1-4\mu -
({\alpha}_0M)^{-2}\}}^{-1}.
\end{eqnarray}
It can be seen that  $A_{eff}\ >\ A$ and so that $A_{eff}$ represents a
'stretched horizon'. In the context of extremal black-holes, the
degeneracy associated with extremal black-hole states is smaller than the
degeneracy of the elementary string states with the same quantum numbers.
To resolve this, Sen [26] postulated that the entropy of the extremal
black-hole is not exactly equal to the area of the event horizon, but the
area of a surface close to the event horizon; the 'stretched horizon'. It
is interesting to see that the introduction of extrinsic curvature action
for strings on the horizon favours this idea of 'stretched' horizon for
neutral Schwarzschild black-hole as well.

\vspace{0.5cm}

\noindent{\bf Acknowledgement}

This work is supported by an operating grant (K.S.V) from the Natural
Sciences and Engineering Research Council of Canada. One of us (R.P)
thanks the Department of Physics, Simon Fraser University, for
hospitality.
\newpage

\vspace{0.5cm}

\noindent{\bf {REFERENCES}}

\begin{enumerate}

\item F.Englert, L.Houart and P.Windey, Phys.Lett. {\bf 372B} (1996) 29.

\item J.D.Bekenstein, Phys.Rev. {\bf D7} (1973) 2333.

\item S.Hawking, Nature. {\bf 248} (1974) 30; Comm.Math.Phys. {\bf 43}
(1975) 199.

\item S.Coleman, J.Preskill and F.Wilczek, Nucl.Phys. {\bf 378B} (1992)
175.

\item F.Dowker, R.Gregory and J.Traschen, Phys.Rev. {\bf D45} (1992) 2762.

\item M.Banados, C.Teitelboim and J.Zaneli, Phys.Rev.Lett. {\bf 72}
(1994) 957. \\
      C.Teitelboim, Phys.Rev. {\bf D51} (1995) 4315. \\
      S.Carlip and C.Teitelboim, Class. Quant. Grav. {\bf 12} (1995) 1699.

\item A.Strominger and C.Vafa, "Microscopic origin of the
Bekenstein-Hawking entropy", \\
      HUTP-96/A002,RU-96-01; hep-th/9601029.

\item G.T.Horowitz, "The origin of black-hole entropy in string theory", \\
      UCSBTH-96-07; gr-qc/9604051.

\item G.T.Horowitz, J.M.Maldacena and A.Strominger, Phys.Lett. \\
      {\bf 383B} (1996) 151.

\item D.M.Kaplan, D.A.Lowe, J.M.Maldacena and A.Strominger, "Microscopic \\
      entropy of N=2 extremal blackholes", CALT-68-2076, RU-96-88, \\
      hep-th/9609204.

\item J.M.Maldacena and A.Strominger, "Statistical entropy of
4-dimensional \\
      extremal blackholes", hep-th/9603060.

\item A.M.Polyakov, Nucl.Phys. {\bf B268} (1986) 406.

\item H.Kleinert, Phys.Lett. {\bf 174B} (1986) 335.

\item T.L.Curtright, G.I.Ghandour and C.K.Zachos, Phys.Rev. {\bf D34}
(1986) 3811.

\item K.S.Viswanathan, R.Parthasarathy and D.Kay, Ann.Phys.(N.Y) {\bf
206} \\
      (1991) 237.

\item R.Parthasarathy and K.S.Viswanathan, Int.J.Mod.Phys. {\bf A7}
(1992) \\
      317;1819.

\item K.S.Viswanathan and R.Parthasarathy, Int.J.Mod.Phys. {\bf A7}
(1992) 5995.

\item K.S.Viswanathan and R.Parthasarathy, Phys.Rev. {\bf D51} (1995) 5830.

\item K.S.Viswanathan and R.Parthasarathy, "String theory in curved
space-time", \\
      SFU-HEP-04-96; hep-th/9605007.

\item D.V.Fursaev, Phys.Rev. {\bf D51} (1995) R5352.

\item G.W.Gibbons and S.Hawking, Phys.Rev. {\bf D15} (1977) 2752.

\item I.Ya.Arefe'va, K.S.Viswanathan and I.V.Volovich, Nucl.Phys. \\
      {\bf B452} (1995) 346; {\bf B462} (1995) 613.

\item S.Hawking ,"The path integral approach to quantum gravity", \\
      in General Relativity: An Einstein Centenary Survey, Eds. S.Hawking
      \\
      and W.Israel, (Cambridge University Press, 1979).

\item T.J.Willmore, Total Curvature in Riemannian Geometry, \\
      Ellis Horwood Ltd. 1982.

\item R.Parthasarathy and K.S.Viswanathan, (to be published).

\item A.Sen, "Extremal blackholes and elementary string states",
hep-th/9504147.

\end{enumerate}
\end{document}